\documentclass[prx,twocolumn,superscriptaddress,showpacs,floatfix,longbibliography]{revtex4-1}
\usepackage{amsmath,amssymb,amsfonts,float,graphics,epsfig,epstopdf,color,verbatim,tabularx,bm,multirow,appendix,hyperref}
\usepackage{lmodern}

\usepackage{color}
\usepackage{bm}

\newcommand{\beq} {\begin{equation}}
\newcommand{\eeq} {\end{equation}}
\newcommand{\bea} {\begin{eqnarray}}
\newcommand{\eea} {\end{eqnarray}}
\newcommand{\be} {\begin{equation}}
\newcommand{\ee} {\end{equation}}
\renewcommand{\(}{\left(}
\renewcommand{\)}{\right)}
\renewcommand{\[}{\left[}
\renewcommand{\]}{\right]}

\DeclareMathOperator{\sgn}{sgn}

\newcommand{\ket}[1]{\left|#1\right>}
\newcommand{\bra}[1]{\left<#1\right|}

\begin{document}

\title {Yukawa-SYK model and Self-tuned Quantum Criticality}
\author{Gaopei Pan}
\affiliation{Beijing National Laboratory for Condensed Matter Physics and Institute of Physics, Chinese Academy of Sciences, Beijing 100190, China}
\affiliation{School of Physical Sciences, University of Chinese Academy of Sciences, Beijing 100190, China}
\author{Wei Wang}
\affiliation{Beijing National Laboratory for Condensed Matter Physics and Institute of Physics, Chinese Academy of Sciences, Beijing 100190, China}
\affiliation{School of Physical Sciences, University of Chinese Academy of Sciences, Beijing 100190, China}
\author{Andrew Davis}
\author{Yuxuan Wang}
\email[]{yuxuan.wang@ufl.edu}
\affiliation{Department of Physics, University of Florida, Gainesville, FL 32601}
\author{Zi Yang Meng}
\email[]{zymeng@hku.hk}
\affiliation{Department of Physics and HKU-UCAS Joint Institute of Theoretical and Computational Physics, The University of Hong Kong, Pokfulam Road, Hong Kong SAR, China}
\affiliation{Beijing National Laboratory for Condensed Matter Physics and Institute of Physics, Chinese Academy of Sciences, Beijing 100190, China}
\affiliation{Songshan Lake Materials Laboratory, Dongguan, Guangdong 523808, China}

\begin{abstract}
Non-Fermi liquids (NFL) are a class of strongly interacting gapless fermionic systems without long-lived quasiparticle excitations. An important group of NFL model features itinerant fermions coupled to soft bosonic fluctuations near a quantum-critical point (QCP), and are widely believed to capture the essential physics of many unconventional
superconductors. However numerically the direct observation of a canonical NFL behavior in such systems, characterized by a power-law form in the Green's function, has been elusive. Here we consider a Sachdev-Ye-Kitaev (SYK)-like model with random Yukawa interaction between critical bosons and fermions (dubbed Yukawa-SYK model). We show it is immune from minus-sign problem and hence can be solved exactly via large-scale quantum Monte Carlo simulation beyond the large-$N$ limit accessible to analytical approaches. Our simulation demonstrates the Yukawa-SYK model features ``self-tuned quantum criticality", namely the system is critical independent of the bosonic bare mass. We put these results to test at finite $N$, and our unbiased numerics reveal clear evidence of these exotic quantum-critical NFL properties -- the power-law behavior in Green's function of fermions and bosons -- which propels the theoretical understanding of critical Planckian metals and unconventional superconductors.
\end{abstract}
\date{\today}
\maketitle

\section{Introduction}
\label{sec:I}
The non-Fermi liquid (NFL) is a state of gapless fermionic matter that does not have long-lived quasiparticles due to its strongly interacting nature~\cite{Stewart2001,Abanov2003}. It is widely believed to be the microscopic origin of the ``strange metal" state observed in a broad range of materials, such as Cu-based ~\cite{Keimer2015} and Fe-based~\cite{ZhaoyuLiu2016,YanhongGu2017} high-temperature superconductors, heavy-fermion compounds~\cite{Custers2003,BinShen2019}, and recently in twisted 2D heterostructures~\cite{YCao2019,shen2019observation}. Additionally, the understanding of the unconventional superconducting phase in these systems naturally hinges on the understanding of the NFL ``normal state". Moreover, recently from the studies of the Sachdev-Ye-Kitaev (SYK) models~\cite{SY,K,SYK2,SYK3}, it has been realized that NFLs host a hidden connection {between strange metals~\cite{guo-gu-sachdev-2020} and} holographic quantum matters that saturate the upper bound for quantum chaos, opening an entirely new avenue in understanding the behavior of NFLs~\cite{YingfeiGu2017}.

The term ``non-Fermi liquid" captures the failure of conventional perturbative approach in treating interacting fermion systems with weak interactions, which poses a challenge in the theoretical understanding of such systems. In a general context, NFL behavior often occurs via electron interactions mediated by gapless bosonic modes~\cite{max,max2,max_last,raghu_15,steve_sam,lawler-barci-fernandez-2006,lawler-fradkin-2007,XiaoYanXu2017,ZiHongLiu2018,ZiHongLiu2019,XiaoYanXu2019,XiaoYanXuReview2019} that render the electrons incoherent. Such gapless bosons typically arise in the vicinity of 
a quantum-critical point (QCP) or in quantum gauge theories. Despite the simplicity of the setup, the analytical solution to these models remains challenging due to the lack of a natural small control parameter. Advancement has been made via modifying the model to a large $N$ limit with $N$ the number of fermion flavors and a leap of faith that the same physics holds down to $N=O(1)$, while these large-$N$ approaches face important subtleties in two spatial dimensions~\cite{sslee}.

Along a separate path, there has been great progress in the numerical front in recent years, in particular in designer Hamiltonian of critical bosons Yukawa-coupled to Fermi surfaces~\cite{Berg2012,Berg2019,XiaoYanXuReview2019}. Recent results in minus-sign-problem-free quantum Monte Carlo (QMC) simulations~\cite{XiaoYanXu2017,ZiHongLiu2018,ZiHongLiu2019,XiaoYanXu2019,XiaoYanXuReview2019,Berg2019} have shown strong evidence of NFL states in a range of such boson-fermion models 
{with gapless bosons from a nematic~\cite{Schattner2016} and ferromagnetic~\cite{XiaoYanXu2017,XiaoYanXu2020} quantum critical points and with gauge fields~\cite{XiaoYanXu2019,ChuangChen2019,ChuangChen2020}}
 (see Ref.~\cite{XiaoYanXuReview2019} for a recent review). It is now possible to obtain accurate and reliable information about the scaling behaviors in the close vicinity of these QCPs, testing and improving our theoretical knowledge about these challenging problems.

To reveal NFL physics in numerics, this class of models require tuning the mass of the boson to a critical value, while away from the QCP the system restores Fermi liquid behavior. However, the precise determination of the quantum critical point and region of NFL are subject to finite size effects, and the position of the QCP is not universal but system dependent. Moreover, recenlty it is found that to reveal the clear signature of NFL in fermion self-energy in these QCP systems, one would also need to control the strength of the effective coupling between fermions and bosons, as well as deduct the nonnegligible thermal contributions to the fermionic self-energy~\cite{XiaoYanXu2020}. These difficulties make it harder to directly reveal the scaling form of the NFL self-energies in these systems.

\begin{figure}[!htp]
	\centering
	\includegraphics[width=0.8\columnwidth]{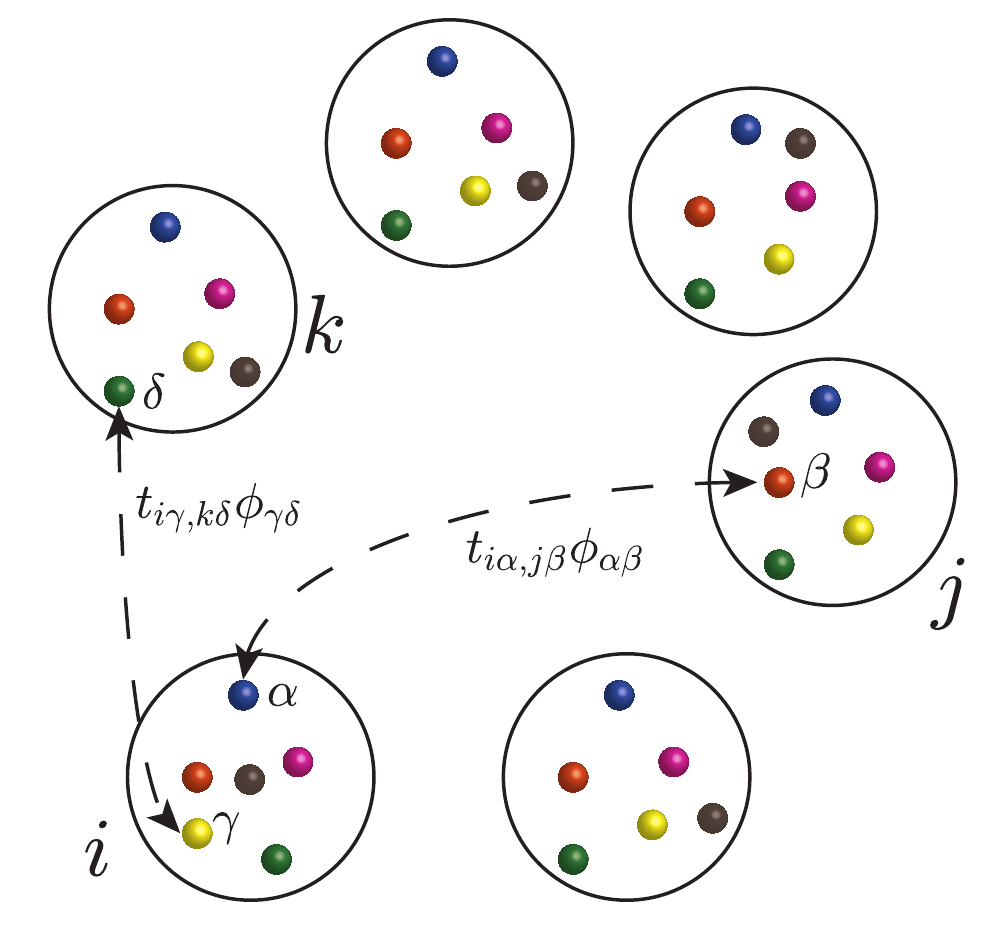}
	\caption{Yukawa-SYK model. There are $M$ quantum dots labeled by $\{i,j\}$, and each dot havs $N$ flavors labeled by $\{\alpha,\beta\}$. Bosons are given by antisymmetric fields $\phi_{\alpha\beta}$. Fermions are coupled to bosons through a  random Yukawa coupling $t_{i\alpha,j\beta }$.} 
	\label{fig:1}
\end{figure}

Recently a class of SYK-like models featuring random Yukawa interactions between bosons and fermions has been put forward to analyze the NFL pairing problem~\cite{wang-2019, esterlis-2019, schmalian-2019,wang-chubukov-2020}. Analytically solvable in a large-$N$ limit similar to the SYK models, the ``Yukawa-SYK model" takes a different approach from the perturbative one to address the interacting fermion system by eliminating kinetic energies from the outset. Physically, such a theoretical approach is of relevance to systems where the Fermi energy is small, e.g., systems with low electron density such as SrTiO$_3$, and Moir\'e flat band systems such as twisted bilayer graphene~\cite{Bistritzer2011,YDLiao2020}. The Yukawa-SYK models have been shown to be maximally chaotic~\cite{kim-cao-altman-2019} and thus likely to admit a holographic dual description. Compared with the SYK models that only involve interacting fermions, the inclusion of a dynamical bosonic degree of freedom in the Yukawa-SYK makes it ideal to model strongly interacting fermionic systems near a QCP. 

Unlike finite dimensional models with quantum-critical points, within large-$N$ approximation these models have been shown to ``self-tune" to quantum criticality, i.e., the system becomes critical due to the strong mutual feedback between the bosonic and fermionic sectors, independent of the bosonic bare mass. In addition, the pairing behavior at large-$N$ has been analytically studied~\cite{wang-2019, esterlis-2019, schmalian-2019}. Depending on details of the Yukawa coupling, these models either show exotic pairing of incoherent fermions, or a NFL phase that is stable to pairing down to $T=0$. While the onset temperature of pairing may be finite, the feedback effects of pairing fluctuations to the fermion Green's function are small ($\sim\mathcal{O}(1/MN)$), without affecting the NFL behavior of the normal state.

As in the original SYK model, these analytical results of the NFL behavior are formally obtained using the replica trick and then taking the replica diagonal ansatz. This is equivalent with replacing the quenched disorder with annealed disorder, which is usually justified by the fact that replica-nondiagonal processes are suppressed by $1/N$~\cite{gu-qi-2016}.
However, the validity of this ansatz is far from obvious~\cite{baldwin-swingle-2019}, since it is not clear whether the summation of the subdominant processes, each small in $1/N$, is convergent. If the system breaks replica symmetry, the true ground state is then a spin glass. For example, replica symmetry breaking occurs in the bosonic SYK model~\cite{gps-2001,fu-sachdev-2016}, and in fact it has been shown recently that similar situations occur for all random interacting bosonic models~\cite{baldwin-swingle-2020}. On the other hand, for the fermionic SYK model, there is now strong numerical and analytical evidence that a glass phase is absent and the NFL state persists down to $T=0$~\cite{gps-2001,fu-sachdev-2016,gur-ari-2018}. For this reason the validity of the large-$N$ analytical result of the Yukawa-SYK model needs to be carefully investigated, especially since the model involves both fermions and bosons. To this end, unbiased numeric calculations, similar in the spirit to the aforementioned critical bosons Yukawa-coupled to Fermi surfaces systems~\cite{Berg2012,Berg2019,XiaoYanXuReview2019,XiaoYanXu2017,ZiHongLiu2018,ZiHongLiu2019,XiaoYanXu2019,Schattner2016,Gerlach2017,ChuangChen2019}, are highly desirable. 

With this motivation in mind, here we address such a timely issue by showing that at finite $N$ the Yukawa-SYK model can be exactly solved by determinantal QMC simulations, thanks to the bosonic degree of freedom. A simple extension of the original model, introducing an antiunitary time-reversal symmetry, eliminates the minus-sign problem without altering the essential physics. {To enable a direct comparison with QMC, we numerically solve the self-energies of the Yukawa-SYK model within large-$N$ at finite temperatures with discrete imaginary time steps. At low temperatures, this indeed agrees with the analytical solution of the Schwinger-Dyson equations with an emergent time reparametrization symmetry. This emergent symmetry indicates that the effect of thermal fluctuations can be easily incorporated in the time domain (see Ref.~\cite{Klein2020,XiaoYanXu2020} for subtleties in the frequency domain), enabling a direct identification of the NFL behavior at finite temperatures.
	
In this paper, we found that as one progressively increase $N$, the Green's functions from QMC simulations do approach the large-$N$ result and display self-tuned criticality and NFL behavior with power-law self-energies. Additionally we found that as $N$ increases, the QMC results with different realizations of the random interaction self-average, i.e., the variance of the Green's function decreases with increasing $N$ and we obtained a good match the large-$N$  results.  This is strong evidence that the system is free from glassy behavior at least within the temperature range accessible to QMC. By comparing with large-$N$ results, we analyze the behavior of finite-$N$ corrections and show that it is consistent with those from replica off-diagonal fluctuations and pairing fluctuations, which decreases with $N$. By contrast, we consider a model~\cite{wang-2019} in a similar form with a crucial difference that the random coupling is of a lower rank. In such a model with less randomness, replica off-diagonal processes are less suppressed. We numerically show that the bosonic Green's function exhibits glassy behavior.

\section{The model}
The model studied here describes $M$ quantum dots each hosting $N$ flavors of fermions interacting with $N^2$ flavors of matrix bosons via all-to-all random Yukawa interactions. The Hamiltonian of this Yukawa-SYK model is given by
\begin{align}
		H =& \sum_{i,j=1}^{M}\sum_{\alpha,\beta=1}^{N} \sum_{m,n}^{\uparrow,\downarrow}\(\frac{i}{\sqrt{MN}} t_{i\alpha,j\beta}\phi_{\alpha\beta}c^\dagger_{i \alpha;m}\sigma^z_{m,n}c_{j\beta ;n} 
		\) \nonumber\\
		&+\sum_{\alpha , \beta =1}^N\(\frac{1}{2}\pi_{\alpha\beta}^2+\frac{m_0^2}{2}\phi_{\alpha\beta}^2\),
		\label{eq:eq1}
	\end{align}
where the random coupling between fermion and boson satisfies $\langle t_{i\alpha,j\beta}\rangle=0$, $\left\langle t_{i \alpha, j \beta} t_{k \gamma, l \delta}\right\rangle=\left(\delta_{\alpha \gamma} \delta_{i k} \delta_{\beta \delta} \delta_{j l}+\delta_{\alpha \delta} \delta_{i l} \delta_{\beta \gamma} \delta_{j k}\right) \omega_{0}^{3}$. This model is very similar to that studied in Ref.~\cite{wang-2019}, the only difference being that here the random coupling $t_{i\alpha,j\beta}$ has a higher rank than that in Ref.~\cite{wang-2019} thta does not depend on $\alpha$ and $\beta$.  As we will see in Sec.~III, the high rank randomness of the Yukawa coupling $t_{i\alpha,j\beta}$ is crucial for stabilizing the non-Fermi liquid behavior. $\pi_{\alpha \beta}$ is the canonical momentum of $\phi_{\alpha \beta}$. Hermiticity of the first term requires $\phi_{\alpha \beta}=-\phi_{\beta \alpha}$. As schematically depicted in Fig.~\ref{fig:1}, here $(\alpha, \beta)$ are flavor indices and $(i,j)$ are site indices. $\sigma^z$ is the Pauli matrix in the fermion spin space for each flavor. In the absence of a chemical potential term $\mu=0$, the model has an exact particle-hole symmetry under $c\to c^\dagger$. The general case with $\mu\neq 0$ has also been recently analytically solved at the $N,M\to \infty$ limit~\cite{wang-chubukov-2020} and will be left for our future numerical studies. Importantly, compared to the model studied in  Ref.~\cite{wang-2019}, this model has a time reversal symmetry $c^\dagger \to c^\dagger i\sigma^y \mathcal{K}$, where $\mathcal{K}$ is the complex conjugation operator, which guarantees the absence of minus-sign problem of the QMC simulation.  For the sake of simplicity we set $\omega_0=1$ as the energy unit throughout the paper. The only other energy scale in Eq.~\eqref{eq:eq1} is the bosonic bare mass $m_0$. We refer to  situations with small and large $\omega_0/m_0$ as ``weakly coupled" and ``strongly coupled".

In the $N\to \infty$, $M\to\infty$ limit, the ground state of the system has been analytically solved~\cite{wang-2019, esterlis-2019}, and the ground state is found to be a non-Fermi liquid. The large-$N$ result is based on the assumption that the replica symmetry of the random model is unbroken. In this work we verify the validity of the non-Fermi liquid solution by examining and extrapolating the system behavior at finite $N,M$. For such $N,M$ analytical calculations are uncontrolled.} Fortunately, due to the time-reversal symmetry in our designer Hamiltonian in Eq.~\eqref{eq:eq1} there is no minus-sign problem (the proof of this is given in Appendix.~\ref{sec:IIIC}).

\subsection{Normal-state results at $N, M\to \infty$}
\label{sec:IIA}
Before demonstrating our QMC results for the Yukawa-SYK model, we first briefly review the theoretical results at the $N, M\to \infty$ limit. In this limit one can show that the effective action has a saddle point given by the Schwinger-Dyson equations
\begin{align}
     \Pi (i\Omega_n) =& \frac{4M}{N}\omega_0^3T\sum_{m}
     G_f(i\omega_m-i\Omega_n/2) G_f(i\omega_m+i\Omega_n/2)\nonumber\\
    \Sigma(i\omega_m) =& -\omega_0^3
T\sum_{m}
     G_b(i\Omega_n) G_f(i\omega_m-i\Omega_n),
     \label{eq:eq2}
\end{align}
where $\Sigma, \Pi$ are fermionic and bosonic self-energies, and $G_f(i\omega_m) =\[
i\omega_m+\Sigma(i\omega_m)\]^{-1}$ and $G_b(i\Omega_n) =  \[\Omega_n^2+ \Pi(i\Omega_m)+m_0^2\]^{-1}$ are fermionic and bosonic Green's functions.

At $T=0$, it was found~\cite{wang-2019,esterlis-2019} that for $m_0\sim\omega_0$ and $\omega,\Omega\ll \omega_0$ the self-energies are given by
\begin{align}
	\Sigma(\omega) =&-G_f(\omega)^{-1}= ic \sgn(\omega) |\omega|^{x}\omega_0^{1-x},\nonumber\\
	\Pi(\Omega) =&G_b(\Omega)^{-1}= -m_0^2 +c^{-2} \alpha(x)|\Omega|^{1-2x}\omega_0^{1+2x},\nonumber\\
	\label{eq:eq3}
\end{align}
where $c$ is a non-universal $O(1)$ constant, and $0<x<1/2$ is determined by
\be
\frac{4M}{N}=\frac{1/x-2}{1+\sec(\pi x)}
\label{eq:eq4}
\ee
and
\be
\alpha(x) = -\frac{\Gamma^2(-x)}{4\pi \Gamma(-2x)}.
\ee
Compared to the results in Ref.~\cite{wang-2019}, Eq.~\eqref{eq:eq4} is different by a factor of 2 because the addition of the spin degree of freedom $m,n=\uparrow/\downarrow$ in the Hamiltonian in Eq.~\eqref{eq:eq1}. In particular, at $M=N$, one finds $x\approx 0.098$, and for $4M=N$, $x\approx 0.231$.

 \begin{figure}[! htp]
	\centering
	\includegraphics[width=\columnwidth]{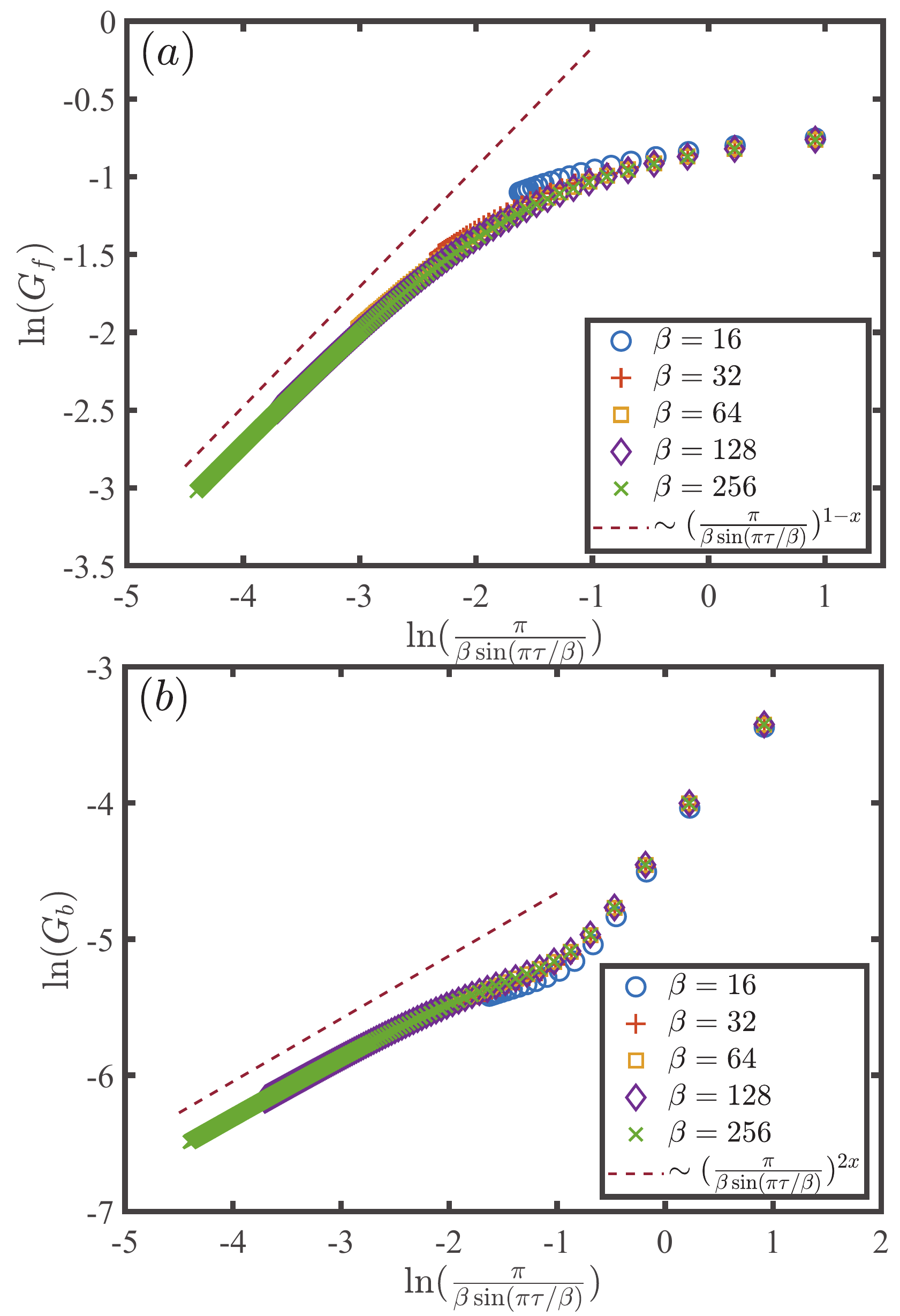}
	\caption{Theoretical result of $G_f$ and $G_b$ at $N=4M\to \infty$, $\omega_0=1, m_0=2$ and various temperatures. Here we show them in log-log plot. The  auxiliary dashed lines whose slopes are $1-x$ and $2x$ show that $G_f(\tau, 0) \propto \(\frac{\pi}{\beta\sin(\pi\tau/\beta)}\)^{1-x}$ and $G_b(\tau,0) \propto \(\frac{\pi}{\beta\sin(\pi\tau/\beta)}\)^{2x}$ at $\tau \rightarrow \frac{\beta}{2}$, when $\beta$ is large enough.}
	\label{fig:2}
\end{figure}

From Eq.~\eqref{eq:eq3} we have
\be
m_0^2-\Pi(\Omega=0)=0,
\ee
indicating that the boson is critical. This was argued in Refs.~\cite{wang-2019, esterlis-2019} to be true for an \emph{arbitrary} $m_0^2$. No matter what the bosonic bare mass is, the system renormalizes it to zero via interaction effects. For this reason we dub this phenomenon ``self-tuned quantum criticality". This feature is certainly not present in any finite dimensional models such as those of critical bosons coupled to Fermi surface systems~\cite{max,max2,max_last,raghu_15,steve_sam,lawler-barci-fernandez-2006,lawler-fradkin-2007,XiaoYanXu2017,ZiHongLiu2018,ZiHongLiu2019,XiaoYanXu2019,XiaoYanXuReview2019,Gerlach2017,XiaoyuWang2017,Berg2019,Schattner2016,ChuangChen2019,XiaoYanXu2020} discussed in Sec.~\ref{sec:I}.

In the time domain, by a Fourier transform we obtain \footnote{The Fourier transform from $\Pi(\Omega)$ is tricky, since the positive power-law $|\Omega|^{1-2x}$ does not have a Fourier transform in the common sense as the Fourier integral is UV divergent for $0<x<1/2$. This divergence is canceled by the Fourier transform of the $m_0^2$ term in $\Pi(\Omega)$ with the UV information.}
\begin{align}
	\Pi(\tau,\tilde\tau) \propto& |\tau-\tilde\tau|^{-(2-2x)},\nonumber\\
	G_b(\tau,\tilde\tau) \propto& |\tau-\tilde\tau|^{-2x},\nonumber\\
	\Sigma (\tau,\tilde\tau) \propto& |\tau-\tilde\tau|^{-(1+x)}\sgn(\tau-\tilde\tau),\nonumber\\
	G_f(\tau,\tilde\tau) \propto& |\tau-\tilde\tau|^{x-1}\sgn(\tau-\tilde\tau).
\end{align}

At a finite temperature $T=1/\beta$, one can accordingly obtain the fermionic and bosonic Green's functions through a reparametrization symmetry transformation$\tau\to f(\tau)=\tan(\pi\tau/\beta)$~\cite{SY,K,SYK2,SYK3} , and we have at low-temperatures and long-time limit,
\begin{align}
	G_f(\tau, 0) \propto & \(\frac{\pi}{\beta\sin(\pi\tau/\beta)}\)^{1-x}\nonumber\\
	G_b(\tau,0) \propto& \(\frac{\pi}{\beta\sin(\pi\tau/\beta)}\)^{2x}.
	\label{eq:eq9}
\end{align}

To enable a direct comparison with the QMC data, we developed an iterative algorithm to solve the nonlinear equation in \eqref{eq:eq2} numerically at an arbitrary temperature. To ensure the convergence of the interations, the temperature dependence of $\Pi(0)$ was fixed using analytical results obtained in Ref.~\onlinecite{esterlis-2019}. As we shall see, the QMC simulations for the bosonic sector are performed on a time lattice with lattice constant $\Delta\tau$, the Matsubara frequencies are compact and defined in a frequency Brillouin zone $\omega_m,\Omega_n  \in (-\pi/\Delta\tau, \pi/\Delta\tau)$. We have incorporated the compactness of the frequency domain within our numerical solution of Eq.~\eqref{eq:eq2} as well, which ensures better match with QMC results especially at large frequencies.

In Fig.~\ref{fig:2} we plot the behavior of $G_f$ and $G_b$ at $N=4M$, $\omega_0=1, m_0=2$  and different temperature from iterative theoretical calculation, in particular, we see that at $\beta=256 \ (T=\frac{1}{256})$, the results matches well in the long-time limit with the approximate result obtained using time-reparametrization symmetry, exhibiting self-tuned criticality and NFL behaviors. This result will later be compared with numerical ones in Sec.~\ref{sec:num}.

\subsection{Pairing at $N, M\to \infty$: Mean field theory}
\label{sec:IIB}

\begin{figure}[!htp]
	\centering
	\includegraphics[width=\columnwidth]{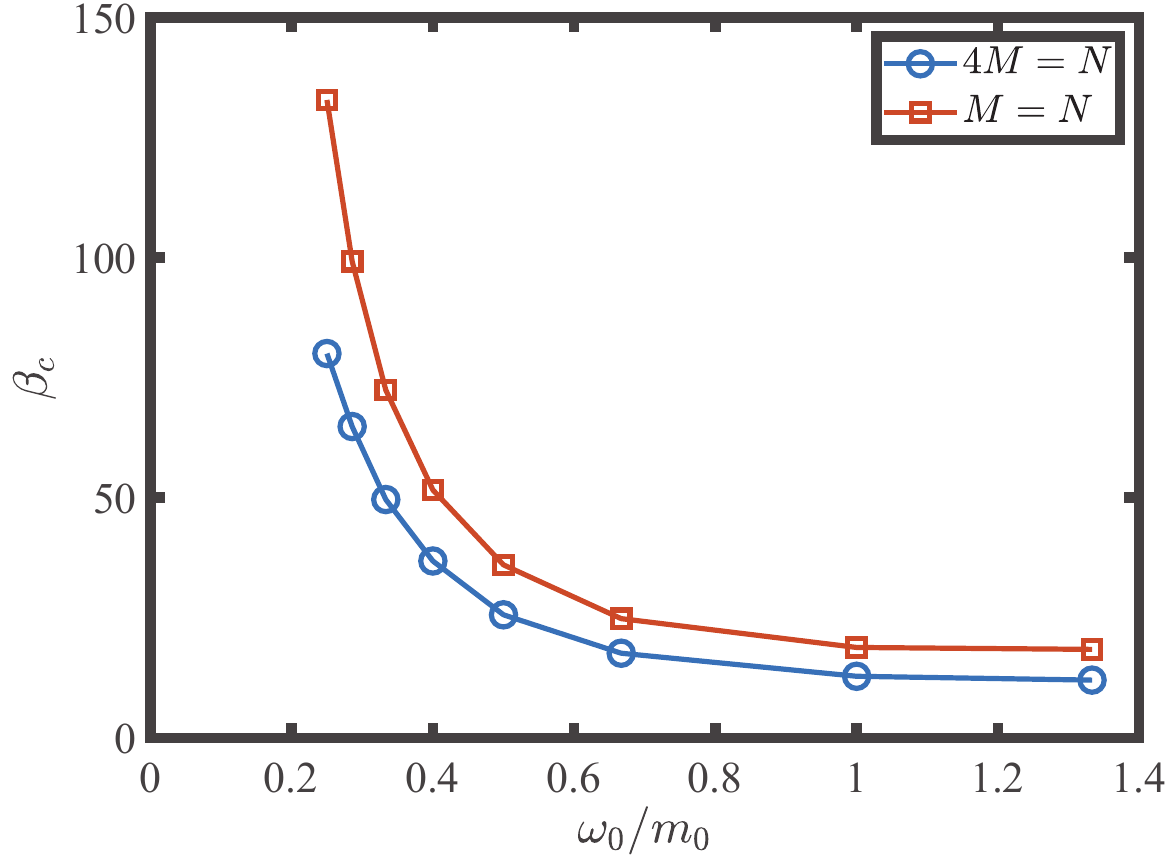}
	\caption{Inverse transition temperature $\beta_c$  from NFL to superconductivity as a function of the ratio $\omega_0/m_0$ for $N=4M$ and $N=M$, obtained from solving Eq.~\eqref{eq:eliash} at large-$N$.}
	\label{fig:3}
\end{figure}

  \begin{figure*}[! htp]
	\centering
	\includegraphics[width=.83\textwidth]{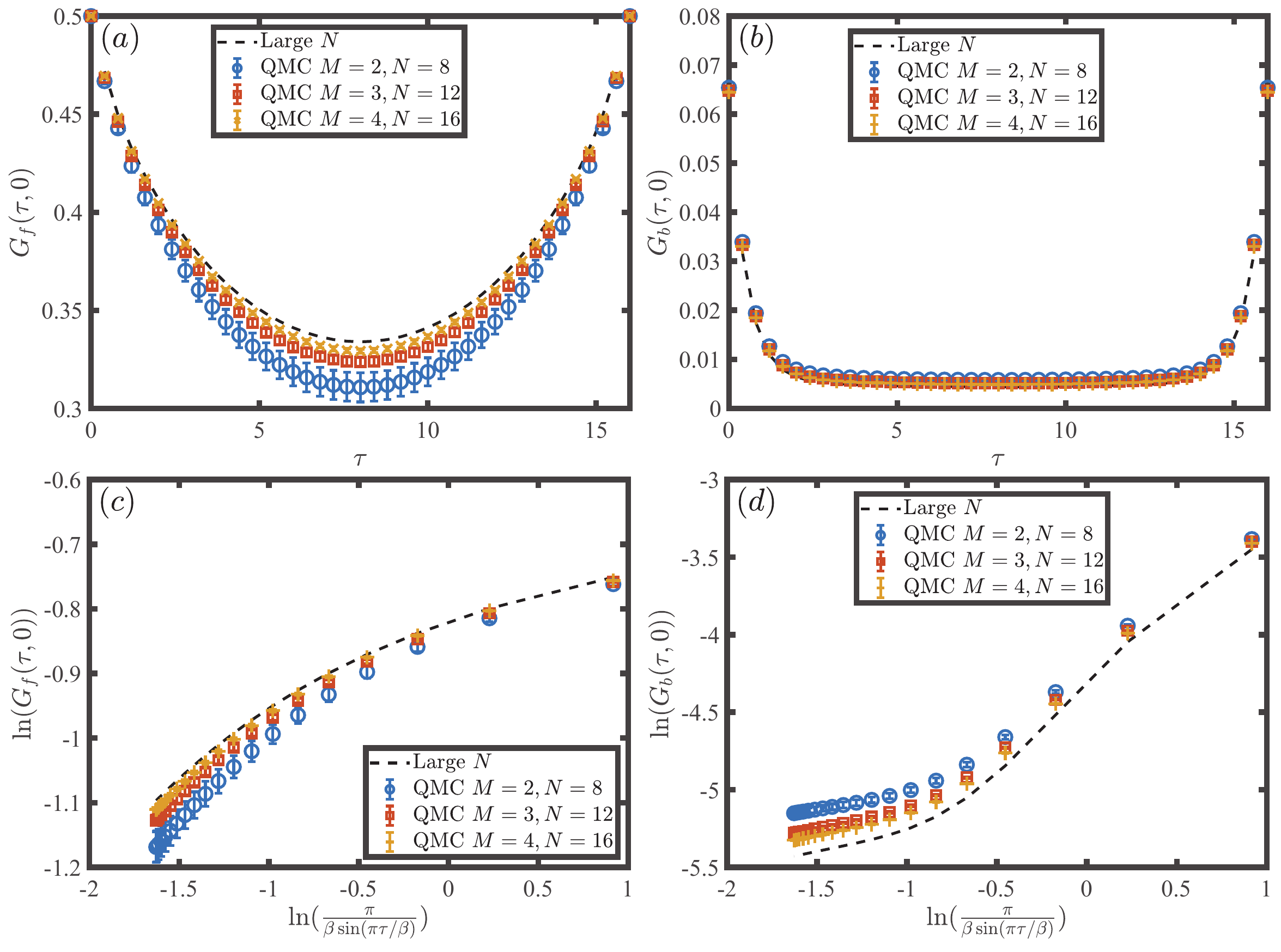}
	\caption{QMC results at $N=4M$, $m_0=2,\omega_0=1$ and $\beta=16$ for $M=2,3,4$. (a) Green's function of fermions $G_{f}(\tau,0)$ versus $\tau$ in the range of $\tau\in[0,\beta]$. Blue, red and yellow dots are DQMC data and the black dashed line is the large-$N$ result. (b) Green's function of bosons $G_{b}(\tau,0)$ versus $\tau$ in the range of $\tau\in[0,\beta]$. (c) and (d) The same as above, but in a special log-log scale as in Fig.~\ref{fig:2}.
	The convergence towards the large-$N$ results as $(M,N)$ increase is obvious. In all panels, the error bar denotes the variation of the Green's function for different disorder realizations. The progressively small error bars as $M$ increases indicate that the randomness of the coupling self-averages.} 
	\label{fig:4}
\end{figure*}

It is straightforward to see that the leading pairing instability mediated by the critical boson mode is toward a spin-singlet, intra-dot, and intra-flavor channel:
\be
\Delta \sim  \sum_{i,\alpha}\langle c^\dagger_{i\alpha\uparrow}c^\dagger_{i\alpha\downarrow}\rangle.
\ee
Within mean-field theory, the pairing behavior is described by the Eliashberg equation
\be
\Delta (i\omega_n) = \omega_0^3  T\sum_{n} G_b(i\Omega_n)|G_f(i\omega_m+i\Omega_n)|^2 \Delta(i\omega_n+i\Omega_m),
\label{eq:eliash}
\ee 
where the $1/MN$ factor given by the two Yukawa interaction vertices has been canceled by the summation of the site and flavor indices of the internal fermions. 

 \begin{figure*}[!htp]
	\centering
	\includegraphics[width=0.83\textwidth]{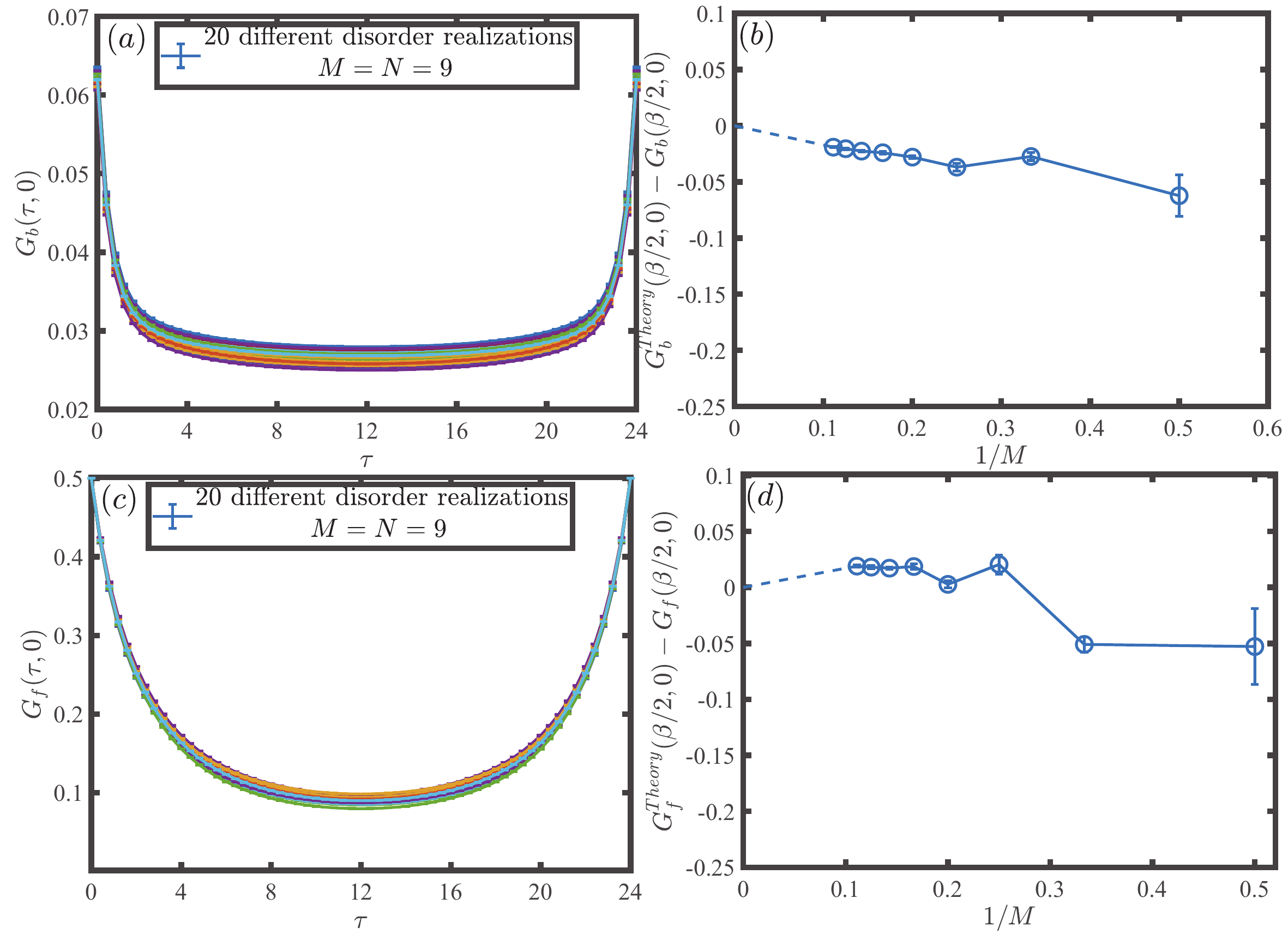}
	\caption{(a) and (c) show $G_b$ and $G_f$ of 20 different disorder realizations with $M=N=9$, $\beta=24$, $m_0=2$ and $\omega_0=1$. The Green's functions are very close to each other. (b) and (d) present the difference between theoretical (large-$N$) results $G^{Theory}(\beta/2,0)$ and QMC numerical simulation data $G(\beta/2,0)$. It is clear that as $M(N)$ increases, the distance between QMC and analytics gradually reduces. And the relative standard deviations in the QMC data are also decreasing. The  parameters are set at $M=N$, $\beta =24$, $m_0=2$ and $\omega_0=1$.}
	\label{fig:5}
\end{figure*}

At $T=0$, plugging in the analytical results in Eq.~\eqref{eq:eq3}, we have
\be
{\Delta(\omega) = \frac{2}{\alpha(x)}\int^{\omega_0}_{\Delta} \frac{d\omega'}{2\pi} \frac{ \Delta(\omega')}{|\omega-\omega'|^{1-2x}|\omega'|^{2x}}.}
\label{eq:eq13}
\ee
where $\Delta$ is the order of magnitude of the frequency-dependent gap $\Delta(\omega)$ that serves as an infrared cutoff of the Green's functions, and $\omega_0\sim m_0$ is an ultraviolet cutoff scale for the low-energy quantum-critical NFL behavior.  

At finite temperatures, we can solve for the critical temperatures $T_c$ using the normal state results numerically obtained. To that end, we numerically solve Eq.~\eqref{eq:eliash} as an eigenvalue problem. As temperature lowers, the eigenvalues of the kernel increases, and $T
_c$ corresponds to the temperature at which the largest eigenvalue approaches one. For reference, we plot $\beta_c$ ($T_c =\frac{1}{\beta_c}$) as a function of the ratio $\omega_0/m_0$ for $N=4M$ and $N=M$ in Fig.~\ref{fig:3}. We see that as  the dimensionless coupling constant $\omega_0/m_0$ increases, $\beta_c$ decreases ($T_c$ increases) in both cases.

At finite $N,M$, a true phase transition to a superconductor does not occur. Yet, pairing fluctuations, which become stronger upon lowering temperatures, do contribute to the fermion Green's function, making the fermions more incoherent. In the Yukawa-SYK model such effects are suppressed by $\mathcal{O}(1/MN)$, but can be detected at small $M,N$. In this sense the fermion Green's function receives two types finite-$N$ corrections -- both from replica-off-diagonal fluctuations and from pairing fluctuations. A true finite temperature phase transition to superconductivity, 
on the other hand, can be obtained by a finite-size extrapolation of the pairing susceptibility  in the QMC simulations. However, the calculation of such observables is beyond the scope of this work and we leave it for future investigations. In this work, we focus on the NFL normal state, although we will discuss signatures of pairing fluctuations in $G_f$ obtained by QMC.
 
 \section{Numerical results}
 \label{sec:num}
 \subsection{NFL Green's functions}
 In this section, we report the key numerical findings in this paper, the NFL Green's function and self-tuned quantum criticality at finite values of $(M,N)$. We choose $\omega_0=1, m_0=2$, such that the dimensionless coupling $\omega_0/m_0$ is reasonably weak, and the pairing fluctuations discussed in Sec.~\ref{sec:IIB} do not significantly modify the normal state NFL behavior.

Fig.~\ref{fig:4} demonstrates the fermion and boson Green's functions obtained in QMC simulation. We focus on $G_b$ and $G_f$ obtained with $N=4M$, $\omega_0=1$, $m_0=2$ at $\beta=16$ for $M=2$, $M=3$ and $M=4$ respectively . Each data point is obtained by averaging over 20 disorder realizations in $\{t_{i\alpha,j\beta}\}$).  Fig.~\ref{fig:4} (a) and (b) are plotted in linear scale, and one can see the QMC curves are progressively close to the large-$N$ curve as $N$ increases. In Fig.~\ref{fig:4} (c) and (d), we present $G_b$ and $G_f$ versus ${\pi}/[{\beta\sin(\pi\tau/\beta)}]$ in a log-log scale, as suggested in Eq.~\eqref{eq:eq9}. It is clear that QMC results match very well with the large-$N$ result, and approach the latter as $N$ increases. The (rather small) error bars in Fig.~\ref{fig:4} denote the variance of the QMC results with different realizations of random couplings.~\footnote{Notice that this is not to be confused with the inherent error of the QMC simulations, whose magnitude is much smaller than the variance from disorder.} We see that such variance decreases upon increasing $N$. This is consistent with the self-averaging behavior of disordered systems and indicates our values of $M=3, N=12$ and $M=4, N=16$ can be reasonably regarded as close to ``large-$N$".

To further quantify the extrapolation to large-$N$, we simulated the model for $N=M$; such a parameter choice allows us to go deeper into the $M$ sequence, and the smaller system size also allows us to go to lower temperatures. In Fig.~\ref{fig:5} (a) and (c), we plot the QMC data of $G_b(\tau,0)$ and $G_f(\tau,0)$ with $M=N=9$, $\beta=24$, $m_0=2$ and $\omega_0=1$, averaged over 20 disorder realizations. As before we see the self-averaging behavior of disorder realizations. In Fig.~\ref{fig:5} (b) and (d), we plot the distance between the QMC disorder averaged Green's functions $G_{b}(\frac{\beta}{2},0)$ and $G_f(\frac{\beta}{2},0)$ and those from the large-$N$ analytical calculation $G^{Theory}_{b}(\frac{\beta}{2},0)$ and $G^{Theory}_{f}(\frac{\beta}{2},0)$. As $1/N\to 0$, indeed $G_b(\frac{\beta}{2},0)$ approaches its large-$N$ value. This indicates the replica-off-diagonal fluctuations are small and suppressed by $1/N$. As a result, glass behavior is absent in this model at least down to $\beta=24$.
 In the meantime, $G_f(\frac{\beta}{2},0)$ is quite close to its large-$N$ value, but remains slightly smaller up to $N=9$. Contrasting the behaviors of $G_b$ and $G_f$, it is tempting to attribute the deviation of $G_f$ to pairing fluctuations. This is consistent with the  
fact that pairing fluctuations makes the fermions more incoherent, and that $T=1/24$ is quite close to the critical temperature with $T_c=1/36$, as shown in Fig.~\ref{fig:3} for the case of $N=M$.  We expect that as $N$ further increases, the effect of pairing fluctuations will be suppressed and the pairing transition at $N=\infty$ is mean-field like. We postpone a detailed study of the pairing transition to future work.

\begin{figure}[!htp]
 	\centering
 	\includegraphics[width=\columnwidth]{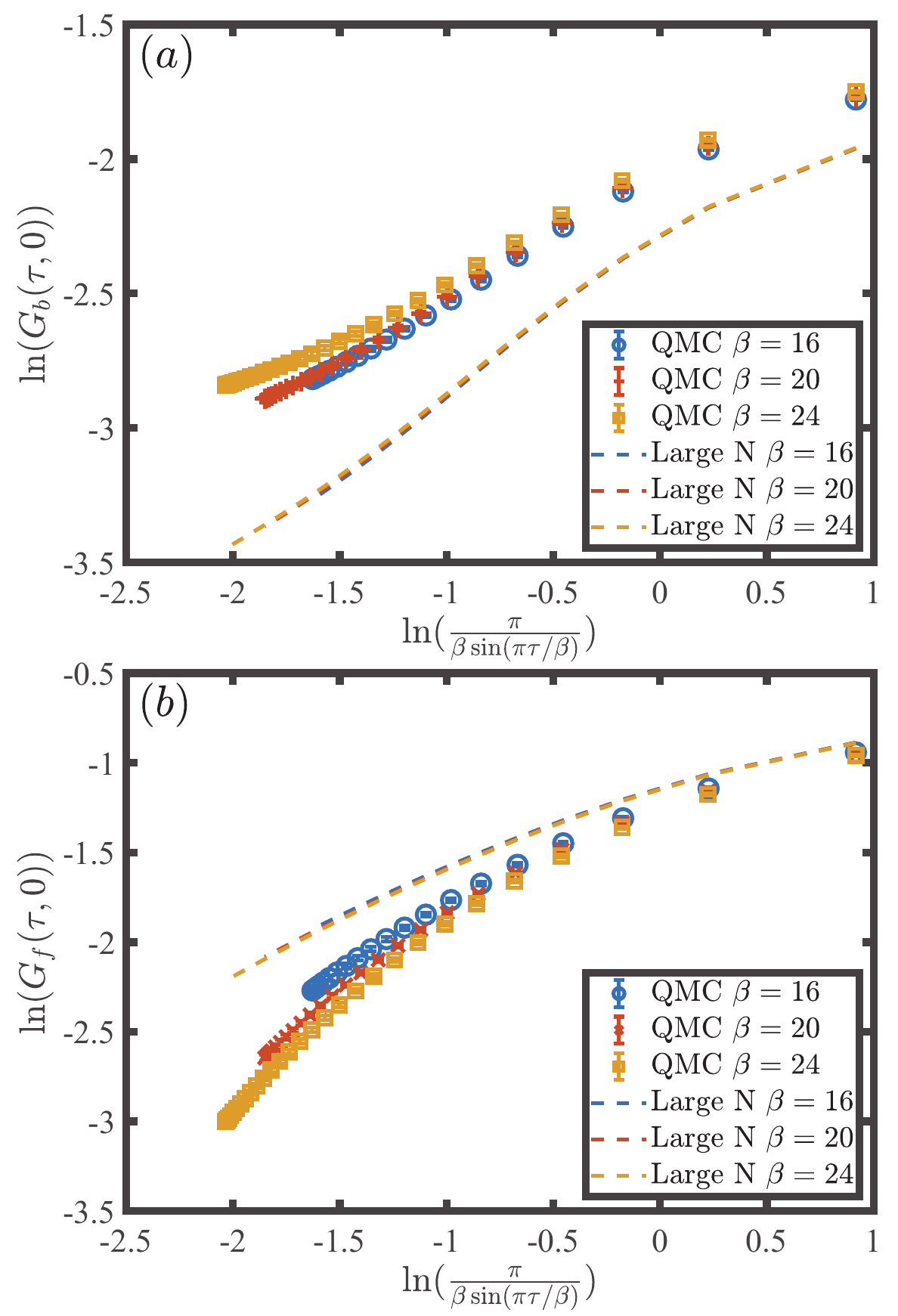}
 	\caption{QMC results of $G_b$ and $G_f$ are at $M=4,N=16$, $m_0=\omega_0=1$, $\beta=16,20,24$. In log-log plot, we see that as the $\beta$ increases, large-$N$ results are basically unchanged, while curves of QMC progressively deviate from the large-$N$ value due to increasing finite-$N$ corrections.}		
 	\label{fig:6}
 \end{figure}
\begin{figure}[!htp]
	\centering
	\includegraphics[width=\columnwidth]{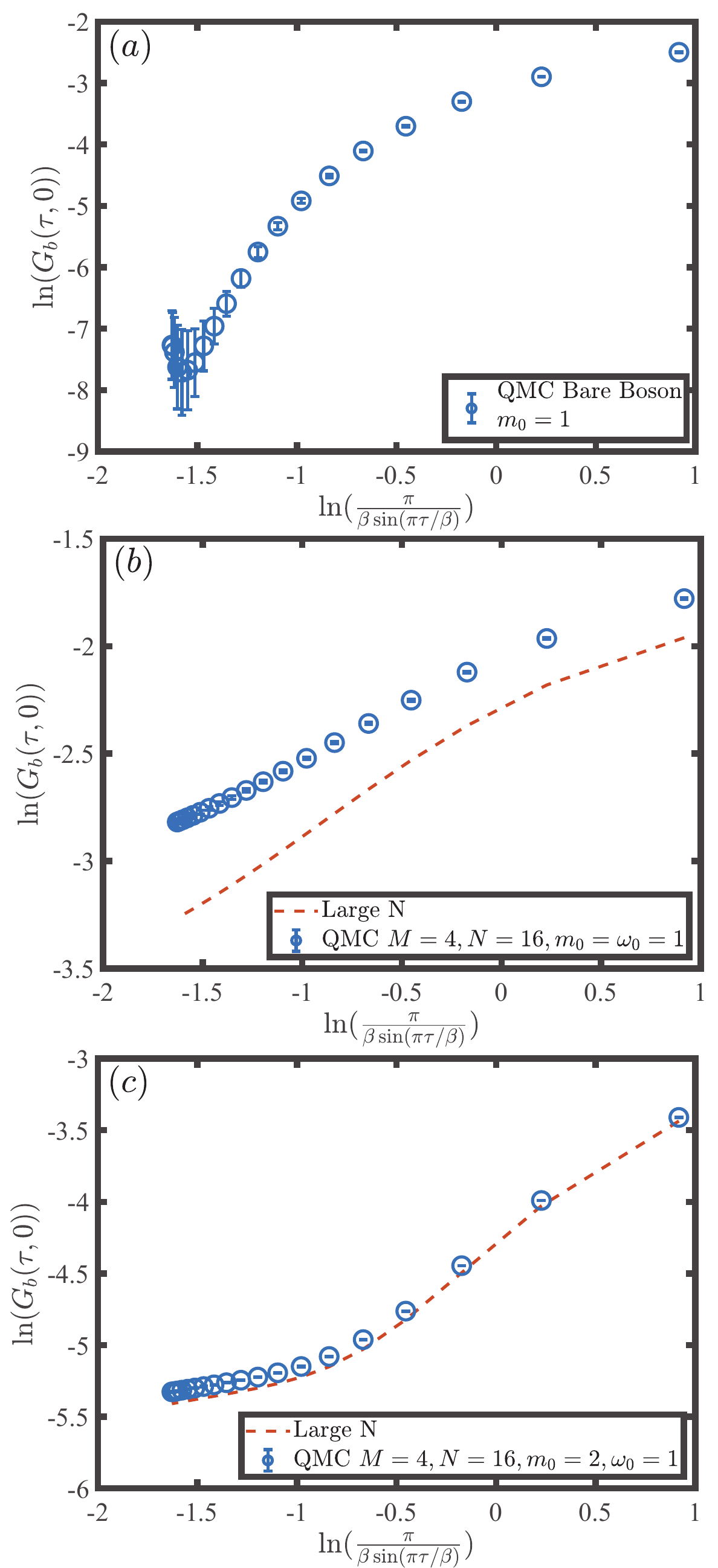}
	\caption{Self-tuned quantum criticality with different boson masses in log-log plot. (a) $G_{b}(\tau,0)$ from a free boson model with $m_0=1$ and $\beta=16$. The exponential decay in imaginary time is evident with $\ln(G_b(\tau=\beta/2)) \sim -7$. (b) and (c) show the $G_{b}(\tau,0)$ from the Yukawa-SYK model in Eq.\eqref{eq:eq1} with different mass $m_0=\omega_0$ [(b)] and $m_0=2\omega_0$ [(c)] with $\omega_0=1$ at $M=4,N=16$ and $\beta=16$. Blue dots are DQMC data and the red dashed lines are large-$N$ result. It is clear to see power law decay of $G_b$ at low-temperatures and long-time limit in log-log plot. These results reveal the self-tuned quantum criticality in our system.}	
	\label{fig:7}
\end{figure}

We emphasize that the randomness of the Yukawa coupling is crucial in stabilizing the NFL behavior. To demonstrate this in Appendix~\ref{sec:appC} we consider a very similar model, in which the random coupling $t_{ij}$ is of lower rank and does not depend on $\alpha, \beta$. This model was analyzed by one of us~\cite{wang-2019} using the Schwinger-Dyson equation at large-$N$, and the analytical results is practically identical to those here. However, our QMC studies have found that its low-temperature phase is actually a spin glass, as the bosonic Green's function has a large static component. Somewhat counterintuitively, the glass phase absent in our present model is realized in such a ``less random" model. Indeed, one can show that in this model, certain replica-off-diagonal diagrams that are not suppressed by $1/N$ survives the disorder averaging, thanks to the lower-rank randomness in the Yukawa coupling, and are expected to drive the glass transition~\cite{wang-unpublished}. Therefore the model studied in Ref.~\cite{wang-2019} needs to be modified.

It is also interesting to investigate the evolution of finite $N$ corrections as a function of temperature. In the original SYK model, it is well known that the strength replica-off-diagonal fluctuations increases with lowering temperature as $\sim 1/N \log^2(T)$, which have led to initial speculations of a glass transition at exponentially low temperatures. On the other hand,  the strength of pairing fluctuations also increases with lowering temperature. To enable a clear analysis of the fluctuation effects, we numerically simulated the intermediate coupling regime of our model with $\omega_0=m_0=1$. Shown in Fig.~\ref{fig:6}, indeed we see that in this case indeed the deviation between numerical and large-$N$ results are more pronounced, and increases upon lowering the temperature. Furthermore, the finite-$N$ corrections modify $G_b$ upward and $G_f$ downward. The upward deviation in $G_b$ is consistent with replica-off-diagonal fluctuations, since they make $G_b$ more static, just as in a glass transition. On the other hand, the downward deviation in $G_f$ is likely to predominantly come from pairing fluctuations.

\subsection{Self-tuned quantum criticality}

As discussed in the Sec.~\ref{sec:IIA}, the self-tuned quantum criticality occurs independently of the bare boson mass $m_0$, at least at the large-$N$ limit. We numerically tested this expectation in QMC simulation with $M=4, N=16$. The results are shown in Fig.~\ref{fig:7}. As a comparison with our interacting model, Fig.~\ref{fig:7} (a) shows the bare boson Green's function generated from $H_b$ in Eq.~\eqref{eq:eq15} and the mass is $m_0=1$ and $\beta=16$. With such a mass term, the Green's function clearly exhibits exponential decay in imaginary time to $G_b(\tau=\beta/2,0)\approx0$. We can see it from the log-log plot: at the far left of the curve $\tau \rightarrow \beta$, which corresponds to the long-time limit, the value of $\ln(G_b(\tau,0))$ decays rapidly. However, once coupled with fermions in our model, as shown in Fig.~\ref{fig:7} (b) [$m_0=\omega_0=1$, averaged over 20 realizations] and (c) [$m_0=2\omega_0=2$, averaged over 20 realizations], with difference masses while keeping the $M=4, N=16$ and $\beta=16$, the boson Green's functions become critical. The Green's functions $G_b$ in imaginary time in  both cases do not decay exponentially, but instead are well consistent with the power-law form of Eq.~\eqref{eq:eq9}.  In (b) and (c), besides the QMC data, we plotted red dashed line which is large-$N$ result. The data in Fig.~\ref{fig:7} (c) turn out to be very close to the theoretical result. Remarkably, here we see that it does not require tuning the bare mass $m_0$ for the system to exhibit quantum-critical behavior, therefore exhibiting the self-tuned quantum criticality, consistent with analytical predictions at large-$N$.

\section{Discussion}
\label{sec:V}
In this work, we  performed unbiased sign-problem-free quantum Monte Carlo simulations of the Yukawa-SYK model, and reported direct evidence of self-tuned quantum-critical and NFL behaviors. We believe such SYK-like models provide a new venue to construct analytical solvable models for strange metals and unconventional superconductors. Our work serves as a starting point of further analyzing such models beyond the analytical large-$N$ limit, in a numerically unbiased manner. Further studies in several further directions are in order.

First, the numerical framework developed here allows one to incorporate the Hubbard $U$ interaction at half-filling without the fermion sign problem. From a theoretical point of view, such a generalized model likely exhibits a strange-metal to Mott insulator transition. It will also be interesting to study if a spin-glass phase can be realized in between, resembling the phase diagram of the underdoped cuprates. Second, recent works have revealed exotic quantum phase transitions between a strange metal and a trivial insulator as one varies the filling~\cite{wang-chubukov-2020}, but  analytical results have only been obtained in the weak-coupling limit. It is an open question whether more exotic phases exist at stronger coupling. Finally, the quantum dot model studied here can be generalized to a lattice model~\cite{XYSong2017,ZhenBi2017,Chowdhury2018}, in which more thermodynamical and transport properties can be examined.

In terms of numerical methodologies, the present work opens the directions of combining the randomness and all-connected models in the study of correlated electron systems, hence greatly broaden the scope of the correlated and itinerant systems. The Yukawa-SYK model and its QMC simulation provide a concrete example of NFL and give us the chance to have a systematic comparison with the large-$N$ analytical calculation. Therefore, one can certian foresee that more realistic and insightful NFL lattice models will eventually be solved with unbiased quantum many-body numerics as the one present here.

\section*{Acknowledgement}
We thank Yingfei Gu, Grigory Tarnopolsky, Subir Sachdev, Steven Kivelson for insightful discussions. G.P.P thanks Rui-Zhen Huang for helpful discussion on numerical calculations. We acknowledge the supports from the Ministry of Science and Technology of China through the National Key Research and Development Program (Grant No. 2016YFA0300502) and Research Grants Council of Hong Kong Special Administrative Region of China through 17303019. We thank the Center for Quantum Simulation Sciences in the Institute of Physics, Chinese Academy of Sciences, the Computational Initiative at the Faculty of Science and the Information Technology Services at the University of Hong Kong, the Platform for Data-Driven Computational Materials Discovery at the Songshan Lake Materials Laboratory, Guangdong, China and  the National Supercomputer Centers in Tianjin and Guangzhou for their technical support and generous allocation of CPU time. This research was initiated at the Aspen Center for Physics, supported by NSF PHY-1066293.

\begin{appendix}
	
	 \section{DQMC methodology}
	 \label{sec:iii}

	 The model described in Eq.~\eqref{eq:eq1} can be solved under the framework of determinant quantum Monte Carlo (DQMC)~\cite{BSS1981,Hirsch1983,Hirsch1985,AssaadEvertz2008,XiaoYanXuReview2019}. DQMC is the method of choice to study the interaction electron systems and has been used extensively in the past few decades in the addressing the problem such as Hubbard~\cite{Hirsch1985}, $t-J$~\cite{Brunner2000} and Kondo lattice~\cite{Assaad1999} models, and lately some great progress have been made in extending the DQMC scheme to interacting topological state of matter~\cite{Hohenadler2012,Meng2014}, duality and QCP beyond Landau-Ginzburg-Wilson paradigm~\cite{YYHe2016,YQQin2017}, and more relevant to this work, the designer Hamiltonians of critical boson coupled to fermions via Yukawa interactions~\cite{Berg2012,XiaoyuWang2017,XiaoYanXu2017,Gerlach2017,ZiHongLiu2018,ZiHongLiu2019,Berg2019,XiaoYanXuReview2019,YuzhiLiu2019,Bauer2020}. In this session, we will elucidate the DQMC setting for model in Eq.~\eqref{eq:eq1} in detail. 
	 
	 First, the partition function reads
	 \begin{eqnarray}  
	 	&Z& =\operatorname{Tr}\left\{e^{-\beta \hat{H}}\right\} =\operatorname{Tr}\left\{\left( e^{-\Delta \tau \hat{H}} \right)^{L_{\tau}}\right\} \nonumber\\
	 	&=& \int (\prod_{\alpha \beta}  \mathrm{d \phi_{\alpha \beta}}) \operatorname{Tr}_{\mathbf{F}} \bra{\phi_{11}\cdots\phi_{NN}}( e^{-\Delta \tau \hat{H}})^{L_{\tau}}\ket{\phi_{11}\cdots\phi_{NN}}\nonumber\\
	 	\label{eq:eq14}
	 \end{eqnarray}
	 where we divide the imaginary time axis into $L_\tau$ slices, $\beta = L_{\tau} \times \Delta \tau$. Let the bosonic configuration at each time slice, $\vec{\Phi}_l=(\phi_{11,l},\phi_{12,l},\cdots,\phi_{N(N-1),l},\phi_{NN,l})$, serves as the complete basis of imaginary time propagation in the path-integral, then
	 \begin{equation}  
	 	\begin{aligned} Z &=\int \left(\prod_{l=1}^{L_{\tau}} \mathrm{d} \vec{\Phi}_l\right)\\&\operatorname{Tr}_{\mathbf{F}} \bra{\vec{\Phi}_1}e^{-\Delta \tau \hat{H}}\ket{\vec{\Phi}_{L_{\tau}}} \bra{\vec{\Phi}_{L_{\tau}}}e^{-\Delta \tau \hat{H}}\ket{\vec{\Phi}_{L_{\tau}-1}}... \\
	 		& \;\; ...\bra{\vec{\Phi}_{2}}e^{-\Delta \tau \hat{H}}\ket{\vec{\Phi}_{1}}. 
	 	\end{aligned}
	 \end{equation}
	 With the help of Suzuki–Trotter decomposition of the Hamiltonian in Eq.~\eqref{eq:eq1}, one has
	 \begin{equation}
	 	e^{-\Delta \tau \hat{H} }\approx e^{-\Delta \tau \hat{H}_{fb} } e^{-\Delta \tau \hat{H}_{b} }
	 \end{equation}
	 where
	 \begin{align}
	 	H_{fb}&=\sum_{i,j}^{M}\sum_{\alpha,\beta}^{N}\sum_{m,n}^{\uparrow\downarrow}\frac{i}{\sqrt{MN}} t_{i\alpha,j\beta}\phi_{\alpha\beta}c^\dagger_{\alpha i m}\sigma^z_{m,n}c_{\beta j n}\label{eq:eq14}\\
	 	H_{b}&=\sum_{\alpha , \beta=1}^N\(\frac{1}{2}\pi_{\alpha\beta}^2+\frac{m_0^2}{2}\phi_{\alpha\beta}^2\),
	 	\label{eq:eq15}
	 \end{align}
	 are the fermion-boson coupled term and the bosonic term, respectively.
	 
	 \subsection{Bosonic Part}
	 Sine we use the space-time arrangement of the bosons $\{\vec{\Phi}_l\}$ to span the configuration space, we need to first express the canonical momentum $\pi_{\alpha\beta}$ in Eq.~\eqref{eq:eq15} in this configuration space. To this end, we first use the coherent state path integral
	 \begin{equation}
	 	\ket{\phi_{\alpha\beta}}=\frac{1}{\sqrt{2 \pi}} \int \mathrm{d \pi_{\alpha\beta}}\; e^{-i \pi_{\alpha\beta} \phi_{\alpha\beta}}\ket{\pi_{\alpha\beta}} 
	 \end{equation}
	 then the momentum term in the partition function can be expressed as
	 \begin{equation} 
	 	\begin{aligned} 
	 		\bra{\phi'} e^{-\frac{1}{2} \Delta\tau \hat{\pi}^2} \ket{\phi}&=\frac{1}{2\pi} \int \mathrm{d \pi}\; e^{i \pi (\phi'-\phi)-\frac{1}{2}\pi^2 \Delta \tau}\\
	 		&\simeq C e^{-\frac{\left( \phi'-\phi\right)^2}{2 \Delta \tau}}
	 	\end{aligned}
	 \end{equation}
	 where $C$ is a constant, and $l$ and $l'$ are two consecutive time slices along the imaginary time axis, and the partition function then becomes
	 \begin{eqnarray} 
	 	&Z&=\int \prod_{l=1}^{L_{\tau}} \mathrm{d} \vec{\Phi}_l \nonumber\\
	 	&&\underbrace{C^{L_{\tau}}( \prod_{l=1}^{L_{\tau}} \prod_{\alpha , \beta=1}^{N} e^{-\Delta \tau \frac{m_0^2}{2}\phi_{\alpha \beta,l}^2})(\prod_{\left<l,l'\right>} \prod_{\alpha , \beta=1}^{N} e^{-\frac{\left( \phi_{\alpha \beta,l}-\phi_{\alpha \beta,l'}\right)^2}{2 \Delta \tau}})}_{\mathcal{W}_b} \nonumber\\
	 	&&\underbrace{\operatorname{Tr}_{\mathbf{F}} \left\{ e^{-\Delta \tau \hat{H}_{fb}\left(\vec{\Phi}_{L_{\tau}}\right)}...\;e^{-\Delta \tau \hat{H}_{fb}\left(\vec{\Phi}_1\right)}\right\}}_{\mathcal{W}_{fb}}
	 	\label{eq:eq8}
	 \end{eqnarray}
	 where the first $\left( \ \right)$ in $\mathcal{W}_b$ contains the spatial boson interaction and the second $\left( \ \right)$ in $\mathcal{W}_b$ contains the temporal boson interaction with $\left<l,l'\right> $ stands for the nearest-neighbor interaction in imaginary time direction, and the $\operatorname{Tr}_{\mathbf{F}}$ in $\mathcal{W}_{fb}$ is the fermion trace we will deal with in Sec.2.B. It is now clear that the Monte Carlo sampling is performed in the bosonic field $\{\vec{\Phi}\}$ space of dimesion $N \times N \times L_{\tau}$ or $MN \times MN \times L_{\tau}$ if one consider the random hopping $t_{\alpha\beta}$ in $H_{fb}$, the configurational weight is comprised of the bosonic part $\mathcal{W}_b$ and the fermion determinant $\mathcal{W}_{fb}$.
	 
	 \subsection{Fermion determinant}
	 \label{sec:IIIB}
	 For a specific bosonic configuration, the fermion determinant is of quadratic form and can be evaluated as that of the free system, following the standard expression
	 \begin{equation}
	 	\operatorname{Tr}_{\mathbf{F}} \left\{ e^{-\sum_{i,j}\hat{c}^\dagger_i A_{i,j}\hat{c}_j}e^{-\sum_{i,j}\hat{c}^\dagger_i B_{i,j}\hat{c}_j}\right\}=\mathrm{Det} \left(\mathbf{I}+e^{-\mathbf{A}}e^{-\mathbf{B}}\right).
	 \end{equation}
	 For the imaginary time propagation in the fermion trace in Eq.~\eqref{eq:eq8}, we define
	 \begin{equation}
	 	B\left(l_2 \Delta \tau, l_1 \Delta \tau \right)=\prod_{l=l_{1}+1}^{l_{2}} e^{-\Delta \tau V(\vec{\Phi}_{l})}
	 	\label{eq:eq23}
	 \end{equation}
	 where
	 \begin{equation} 
	 	\begin{aligned} 
	 		V(\vec{\Phi}_{l})&=\frac{i}{\sqrt{MN}}  \sigma^z_{2\times 2} \otimes \left( t_{i\alpha,j\beta}\phi_{\alpha \beta,l}\right)_{MN \times MN} .
	 	\end{aligned}
	 	\label{eq:eq11}
	 \end{equation}
	 It is interesting to note that in the conventional Hubbard-type model setting, there also exists a fermion hopping matrix on the exponential form, but since here we only have fermion Yukawa coupled with the bosonic field, that the hopping matrix is reduced to identical matrix, and the interaction matrix $V(\vec{\Phi}_l)$, which depends on the space-time configuration of the bosonic field $\{\vec{\Phi}_l\}$, contains both the randomness in hopping matrix $ \sigma^z_{2\times 2} \otimes (t_{i,j})_{M\times M}$ and the bosonic fluctuation matrix $(\phi_{\alpha \beta,l})_{N\times N}$. Such that after tracing out the fermion operators $c^{\dagger}_{\alpha i m}$ and $c_{\beta j n}$, the resulting fermion determinant is the determinant of matrices with size $MN \times MN$ and block diagonal in the fermion spin space of $\sigma^{z}$. 
	 
	 With these notations prepared, finally the partition function in Eq.~\eqref{eq:eq8} can now be written as
	 \begin{eqnarray} 
	 	&Z&=\int \prod_{l=1}^{L_{\tau}} \mathrm{d} \vec{\Phi}_l \nonumber\\
	 	&&\underbrace{C^{L_{\tau}}(\prod_{l=1}^{L_{\tau}} \prod_{\alpha , \beta=1}^{N} e^{-\Delta \tau \frac{m_0^2}{2}\phi_{\alpha \beta,l}^2})(\prod_{\left<l,l'\right>} \prod_{\alpha , \beta=1}^{N} e^{-\frac{\left( \phi_{\alpha \beta,l}-\phi_{\alpha \beta,l'}\right)^2}{2 \Delta \tau}})}_{\mathcal{W}_{b}}\nonumber\\
	 	&&\underbrace{\operatorname{Det}[\mathbf{1}+B(L_\tau \Delta \tau,\(L_\tau-1\)\Delta \tau)\cdots B(\Delta \tau,0)]}_{\mathcal{W}_{fb}} 
	 	\label{eq:eq25}
	 \end{eqnarray}
	 This is the partition function describing the SYK-Yukawa model in Eq.~\eqref{eq:eq1} and we can now simulate it in DQMC.
	 
	 \subsection{Free from sign problem}
	 \label{sec:IIIC}
	 As aforementioned, the partition function in Eq.~\eqref{eq:eq25} is free from the minus-sign problem in the protection of a time-reversal symmetry~\cite{CJWu2005}, i.e., the Hamiltonian is invariant under such a symmetry operation, this can be easily demonstrated as follows. 
	 
	 First, we note 
	 \begin{equation} 
	 	\begin{aligned} 
	 		H_{fb}=\sum_{i,j=1}^{M}\sum_{\alpha,\beta=1}^{N}&\frac{i}{\sqrt{MN}} t_{i\alpha,j\beta}\phi_{\alpha \beta}c^\dagger_{\alpha i \uparrow}c_{\beta j \uparrow} \\
	 		-&\frac{i}{\sqrt{MN}} t_{i\alpha,j\beta}\phi_{\alpha \beta}c^\dagger_{\alpha i \downarrow}c_{\beta j \downarrow}
	 	\end{aligned}
	 \end{equation}
	 and time-reversal symmetry operator is $\mathcal{T}=i\sigma_y \mathcal{K}$. Its operation works as $\mathcal{T} c_{m} \mathcal{T}^{-1} = U_{mn} c_{n} $ , $\mathcal{T} c_{m}^\dagger \mathcal{T}^{-1} = U^{*}_{mn} c_{n}^\dagger  $ , $\mathcal{T} i \mathcal{T}^{-1} = -i $, where $m,n=\uparrow / \downarrow$, $U=i\sigma_y$, then
	 \begin{equation} 
	 	\begin{aligned}
	 		\mathcal{T} H_{fb} \mathcal{T}^{-1}&=\sum_{i,j=1}^{M}\sum_{\alpha,\beta=1}^{N}-\frac{i}{\sqrt{MN}} t_{i\alpha,j\beta}\phi_{\alpha \beta}c^\dagger_{\alpha i \downarrow}c_{\beta j \downarrow}\\
	 		&\qquad  \qquad+\frac{i}{\sqrt{MN}} t_{i\alpha,j\beta}\phi_{\alpha \beta}c^\dagger_{\alpha i \uparrow}c_{\beta j \uparrow} \\
	 		&=H_{fb},
	 	\end{aligned}
	 \end{equation}
	 therefore $H_{fb}$ is invariant under $\mathcal{T}$. 
	 
	 Next, notice that $V\(\vec{\Phi}_l\)$ is block diagonal in the space of $m,n=\uparrow,\downarrow$, then the fermion determinant can be written as
	 \begin{equation} 
	 	\begin{aligned}
	 		\operatorname{Det}&[\mathbf{1}+B(\beta, 0)]\\
	 		&=\operatorname{Det}[\mathbf{1}+B^{\uparrow}(\beta, 0)] \operatorname{Det}[\mathbf{1}+B^{\downarrow}(\beta, 0)]\\
	 		&=\operatorname{Det}[\mathbf{1}+B^{\uparrow}(\beta, 0)] \operatorname{Det}[\mathcal{T}\(\mathbf{1}+B^{\downarrow}(\beta, 0)\)\mathcal{T}^{-1}]^{*}\\
	 		&=\operatorname{Det}[\mathbf{1}+B^{\uparrow}(\beta, 0)] \operatorname{Det}[\mathbf{1}+B^{\uparrow}(\beta, 0)]^{*}\\
	 		&=\left | \operatorname{Det}[\mathbf{1}+B^{\uparrow}(\beta, 0)] \right |^2
	 	\end{aligned}
	 \end{equation}
	 and it is positive definite. Also note that the boson weight $\mathcal{W}_b$ is positive definite as the $\{\vec{\Phi}\}$ is the eigenstate of the $H_b$ in the space-time. So the entire configurational weight is positive definite and there is no sign-problem for the simulation.

	 \subsection{Update and measurement}
	 Another important ingredient in any Monte Carlo simulation is the update scheme between configurations, here since the bosonic fields are continuous variables, we have to adapt to local update with Metropolis-type acceptance rate. 
	 
	 The ensemble average of physical observable can be expressed as:
	 \begin{equation}
	 	\langle\hat{O}\rangle=\frac{\operatorname{Tr}\left\{e^{-\beta \hat{H}} \hat{O}\right\}}{\operatorname{Tr}\left\{e^{-\beta \hat{H}}\right\}}=\int \left(\prod_{l=1}^{L_{\tau}} \mathrm{d} \vec{\Phi}_l\right) \mathcal{P}_{\mathcal{C}}\langle\hat{O}\rangle_{\mathcal{C}}+O\left(\Delta \tau^{2}\right)
	 \end{equation}
	 where $\Delta\tau^2$ systematical error comes from the Trotter decomposition and the weight and expectation value for each bosonic field configuration $\mathcal{C}$ are
	 \begin{eqnarray} 
	 	\mathcal{P}_{\mathcal{C}}&=& \frac{\mathcal{W}_{\mathcal{C}}^{fb} \operatorname{Det}[1+B_{\mathcal{C}}(\beta, 0)]}{\int \left(\prod_{l=1}^{L_{\tau}} \mathrm{d} \vec{\Phi}_l\right) \mathcal{W}_{\mathcal{C}}^{fb} \operatorname{Det}[1+B_{\mathcal{C}}(\beta, 0)]} \\\langle\hat{O}\rangle_{\mathcal{C}}&=& \frac{\operatorname{Tr}\{\hat{U}_{\mathcal{C}}(\beta, \tau) \hat{O} \hat{U}_{\mathcal{C}}(\tau, 0)\}}{\operatorname{Tr}\{\hat{U}_{\mathcal{C}}(\beta, 0)\}},
	 \end{eqnarray}
	 where
	 \begin{equation}
	 	\hat{U}\left(l_2 \Delta \tau, l_1 \Delta \tau \right)=\prod_{l=l_{1}+1}^{l_{2}} e^{-\Delta \tau \mathbf{\hat{c}^{\dagger}} V(\vec{\Phi}_{l}) \mathbf{\hat{c}}}
	 	\label{eq:eq32}
	 \end{equation}
	 here $\mathbf{\hat{c}}$ has $2\times M \times N $ components, so does the dimension of the matrix $V$. Once tracing out the quadratic fermions $\mathbf{\hat{c}}$ in Eq.~\eqref{eq:eq32}, one arrives at the $B(l_2\Delta\tau, l_1\Delta\tau)$ matrix in Eq.~\eqref{eq:eq23}, and the evaluation of fermion determinant follows from there down to Eq.~\eqref{eq:eq25}. The detailed derivation of physical observables, exemplified by the equal time and imaginary time displaced fermionic Green's functions, are given in Supplementary Information (SI).
	 
	 Moreover, since the coupling matrix $t_{ij}$ in $H_{fb}$ is subject to randomness, the aforementioned Monte Carlo sample is performed for each disorder realization. Therefore, besides the Monte Carlo average over a fixed disorder configuration, the final physical observables such as the fermion and boson Green's functions are the disordered averaged quantities.
	
	\section{Monte Carlo measurements}
	\label{app:appA}

	The ensemble average of physical observables, in the DQMC formalism, can be calculated as,
	\begin{eqnarray}
		\langle\hat{O}\rangle_{\mathcal{C}} &=&\left.\frac{\partial}{\partial \eta} \ln \operatorname{Tr}\left[\hat{U}_{\mathcal{C}}(\beta, \tau) e^{\eta \hat{O}} \hat{U}_{\mathcal{C}}(\tau, 0)\right]\right|_{\eta=0} \nonumber\\
		&=&\left.\frac{\partial}{\partial \eta} \ln \operatorname{Det}\left[\mathbf{1}+B_{\mathcal{C}}(\beta, \tau) e^{\eta O} B_{\mathcal{C}}(\tau, 0)\right]\right|_{\eta=0} \nonumber\\
		&=&\left.\frac{\partial}{\partial \eta} \operatorname{Tr}\ln \left[\mathbf{1}+B_{\mathcal{C}}(\beta, \tau) e^{\eta O} B_{\mathcal{C}}(\tau, 0)\right]\right|_{\eta=0} \nonumber\\
		&=&\operatorname{Tr}\left[B_{\mathcal{C}}(\tau, 0)\left(1+B_{\mathcal{C}}(\beta, 0)\right)^{-1} B_{\mathcal{C}}(\beta, \tau) O\right] \nonumber\\ &=&\operatorname{Tr}\left[\left(1-(1+B_{\mathcal{C}}(\tau, 0) B_{\mathcal{C}}(\beta, \tau))^{-1}\right) O\right] 
		\label{eq:eqA1}
	\end{eqnarray}
	in the case of equal time fermionic Green's function, $\hat{O}=\mathbf{\hat{c}}^{\dagger} O \mathbf{\hat{c}}$. $\hat{U}_{\mathcal{C}}$ and $B_{\mathcal{C}}$ are defined in Eq.(23) and Eq.(32) in the main text, respectively.
	
	For the imaginary time displaced fermionic Green's function, $G_{f,ij}(\tau,0)= \langle c_i(\tau) c_j^\dagger (0) \rangle $ where $i,j$ encapsulate the dot, flavor and spin indices in the Hamiltonian in Eq.(1) in the main text and the imaginary time $\tau\in[0,\beta]$, it can be evaluated in DQMC as
	\begin{eqnarray} 
		\langle c_i(\tau) c_j^\dagger (0) \rangle&=&\frac{\operatorname{Tr}\{\hat{U}_{\mathcal{C}}(\beta, \tau)\; \hat{c}_i\; \hat{U}_{\mathcal{C}}(\tau, 0)\hat{c}_j^\dagger\}}{\operatorname{Tr}\{\hat{U}_{\mathcal{C}}(\beta, 0)\}}\nonumber\\
		&=&\frac{\operatorname{Tr}\{\hat{U}_{\mathcal{C}}(\beta, 0)\;[\hat{U}^{-1}_{\mathcal{C}}(\tau, 0) \hat{c}_i\; \hat{U}_{\mathcal{C}}(\tau, 0)]\hat{c}_j^\dagger\}}{\operatorname{Tr}\{\hat{U}_{\mathcal{C}}(\beta, 0)\}} \nonumber\\
		&=&\sum_k B_{\mathcal{C}}(\tau, 0)_{ik} \frac{\operatorname{Tr}\{\hat{U}_{\mathcal{C}}(\beta, 0)\; \hat{c}_k \hat{c}_j^\dagger\}}{\operatorname{Tr}\{\hat{U}_{\mathcal{C}}(\beta, 0)\}} \nonumber\\
		&=&[B_{\mathcal{C}}(\tau, 0)(1+B_{\mathcal{C}}(\beta, 0))^{-1}]_{ij}
		\label{eq:eqA2}
	\end{eqnarray}
	where the intermediate steps in Eq.~\eqref{eq:eqA2} are given explicitly in Ref.~\cite{Hirsch1985}.

		   \begin{figure}[!htp]
	   	\centering
	   	\includegraphics[width=\columnwidth]{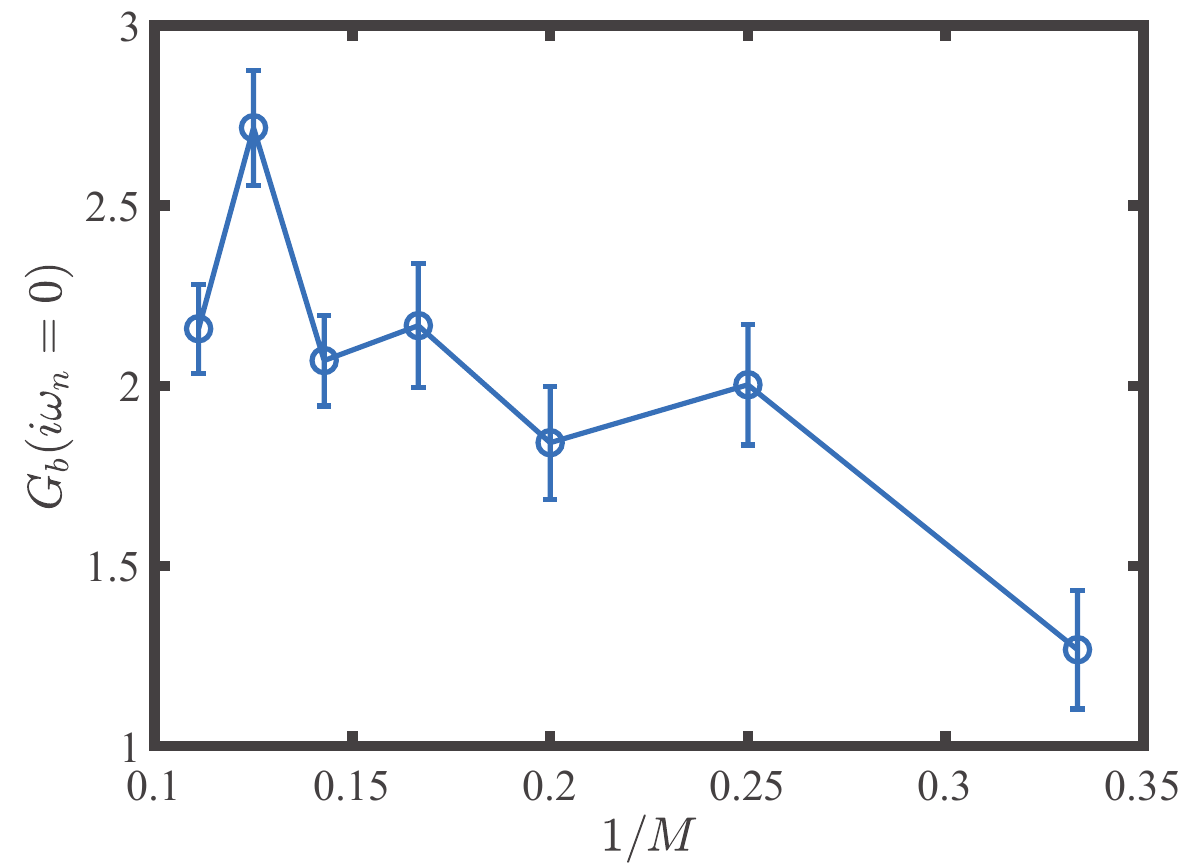}
	   	\caption{DQMC results of Matsubara Green function $G_b$ at $M=N$, $m_0=\omega_0=1$, $\beta=16$. We plot the zero frequency component and its errorbar. Galssy behaviors are seen.}		
	   	\label{fig:8}
	   \end{figure}
	\section{Glass behavior in a less-random model}
	\label{sec:appC}
	We construct a model in a similar form in which the random coupling is of a lower rank,
	 \begin{equation}
	 	\begin{aligned}
	 		H =& \sum_{i,j=1}^{M}\sum_{\alpha,\beta=1}^{N} \sum_{m,n}^{\uparrow,\downarrow}\(\frac{i}{\sqrt{MN}} t_{i,j}\phi_{\alpha\beta}c^\dagger_{i \alpha;m}\sigma^z_{m,n}c_{j\beta ;n} 
	 		\) \\
	 		&+\sum_{\alpha , \beta =1}^N\(\frac{1}{2}\pi_{\alpha\beta}^2+\frac{m_0^2}{2}\phi_{\alpha\beta}^2\),
	 	\end{aligned}
	 \end{equation}
where the random coupling between fermion and boson is realized as $\left\langle t_{i j}\right\rangle=0,\left\langle t_{i j} t_{k l}\right\rangle=\left(\delta_{i k} \delta_{j l}+\delta_{i l} \delta_{j k}\right) \omega_{0}^{3},$ . Still for the sake of simplicity we set $\omega_{0}=1$ as the energy unit throughout the paper, and the temperature scale is then $T \equiv \omega_{0} / \beta . \pi_{\alpha \beta}$ is the canonical momentum of $\phi_{\alpha \beta}$. Hermiticity of the first term requires $\phi_{\alpha \beta}=-\phi_{\beta \alpha}$.

{In Fig.~\ref{fig:8} , we show the static component (with $\omega_n=0$) for bosonic Green's function  $G_b(\omega_n)$. In the large-$N$ limit, this component can be regarded as an Edwards-Anderson order parameter of the spin glass phase~\cite{fu-sachdev-2016}. As $N$ increases, the static component, along with its variance for different disorder realizations,  increases with the increase of $N$ at $M=N, \beta=16, m_0=\omega_0=1$, which is indicative of a spin glass behavior. }

\end{appendix}
	
\bibliographystyle{apsrev4-1}
\bibliography{notes-Yukawa-SYK.bib}
\end{document}